\documentstyle[onecolumn]{mn}

\newif\ifAMStwofonts
\DeclareMathAlphabet\EuFrak{U}{euf}{m}{n}
\SetMathAlphabet\EuFrak{bold}{U}{euf}{b}{n}
  \DeclareFontFamily{U}{euf}{}%
  \DeclareFontShape{U}{euf}{m}{n}{<-6>eufm5<6-8>eufm7<8->eufm10}{}%
  \DeclareFontShape{U}{euf}{b}{n}{<-6>eufb5<6-8>eufb7<8->eufb10}{}%
\newcommand{\mathfrak}{\EuFrak}
\RequirePackage{keyval,graphics}
\RequirePackage{subfigure}
\RequirePackage{graphicx}
\RequirePackage{graphics,psfrag}
\DeclareSymbolFont{AMSa}{U}{msa}{m}{n}
\DeclareSymbolFont{AMSb}{U}{msb}{m}{n}
\DeclareSymbolFontAlphabet{\mathbb}{AMSb}

\ifoldfss
  \ifCUPmtlplainloaded \else
    \NewTextAlphabet{textbfit} {cmbxti10} {}
    \NewTextAlphabet{textbfss} {cmssbx10} {}
    \NewMathAlphabet{mathbfit} {cmbxti10} {} % for math mode
    \NewMathAlphabet{mathbfss} {cmssbx10} {} %  "   "    "
  \fi
  \ifAMStwofonts
    \ifCUPmtlplainloaded \else
      \NewSymbolFont{upmath} {eurm10}
      \NewSymbolFont{AMSa} {msam10}
      \NewMathSymbol{\upi}     {0}{upmath}{19}
      \NewMathSymbol{\umu}     {0}{upmath}{16}
      \NewMathSymbol{\upartial}{0}{upmath}{40}
      \NewMathSymbol{\leqslant}{3}{AMSa}{36}
      \NewMathSymbol{\geqslant}{3}{AMSa}{3E}

      \let\leq=\leqslant 
      \let\geq=\geqslant 
    \fi
  \fi
\fi % End of OFSS

\ifnfssone
  \newmathalphabet{\mathit}
  \addtoversion{normal}{\mathit}{cmr}{m}{it}
  \addtoversion{bold}{\mathit}{cmr}{bx}{it}
  \newmathalphabet{\mathbfit} % math mode version of \textbfit{..}
  \addtoversion{normal}{\mathbfit}{cmr}{bx}{it}
  \addtoversion{bold}{\mathbfit}{cmr}{bx}{it}
  \newmathalphabet{\mathbfss} % math mode version of \textbfss{..}
  \addtoversion{normal}{\mathbfss}{cmss}{bx}{n}
  \addtoversion{bold}{\mathbfss}{cmss}{bx}{n}
  \ifAMStwofonts
    \ifCUPmtlplainloaded \else
      \UseAMStwoboldmath
      \makeatletter
      \new@mathgroup\upmath@group
      \define@mathgroup\mv@normal\upmath@group{eur}{m}{n}
      \define@mathgroup\mv@bold\upmath@group{eur}{b}{n}
      \edef\UPM{\hexnumber\upmath@group}
      \new@mathgroup\amsa@group
      \define@mathgroup\mv@normal\amsa@group{msa}{m}{n}
      \define@mathgroup\mv@bold\amsa@group{msa}{m}{n}
      \edef\AMSa{\hexnumber\amsa@group}
      \makeatother
      \mathchardef\upi="0\UPM19
      \mathchardef\umu="0\UPM16
      \mathchardef\upartial="0\UPM40
      \mathchardef\leqslant="3\AMSa36
      \mathchardef\geqslant="3\AMSa3E

      \let\leq=\leqslant 
      \let\geq=\geqslant 
    \fi
  \fi
\fi % End of NFSS release 1

\ifnfsstwo
  \DeclareMathAlphabet{\mathbfit}{OT1}{cmr}{bx}{it}
  \SetMathAlphabet\mathbfit{bold}{OT1}{cmr}{bx}{it}
  \DeclareMathAlphabet{\mathbfss}{OT1}{cmss}{bx}{n}
  \SetMathAlphabet\mathbfss{bold}{OT1}{cmss}{bx}{n}
  \ifAMStwofonts
    \ifCUPmtlplainloaded \else
      \DeclareSymbolFont{UPM}{U}{eur}{m}{n}
      \SetSymbolFont{UPM}{bold}{U}{eur}{b}{n}
      \DeclareSymbolFont{AMSa}{U}{msa}{m}{n}
      \DeclareMathSymbol{\upi}{0}{UPM}{"19}
      \DeclareMathSymbol{\umu}{0}{UPM}{"16}
      \DeclareMathSymbol{\upartial}{0}{UPM}{"40}
      \DeclareMathSymbol{\leqslant}{3}{AMSa}{"36}
      \DeclareMathSymbol{\geqslant}{3}{AMSa}{"3E}

      \let\leq=\leqslant 
      \let\geq=\geqslant 
    \fi
  \fi
\fi % End of NFSS release 2

\ifCUPmtlplainloaded \else
  \ifAMStwofonts \else % If no AMS fonts
    \def\upi{\pi}
    \def\umu{\mu}
    \def\upartial{\partial}
  \fi
\fi

\title[Two-integral DFs for axisymmetric stellar systems]
  {Two-integral distribution functions for axisymmetric stellar systems with separable densities}
\author[Z.~Jiang and L.~Ossipkov]
  {Zhenglu Jiang$^1$\thanks{mcsjzl@mail.sysu.edu.cn} 
       and Leonid Ossipkov$^2$\thanks{leo@dyna.astro.spbu.ru}
\\
  $^1$Department of Mathematics, Zhongshan University, 
              Guangzhou 510275, China\\
  $^2$Saint~Petersburg State University, Staryj Peterhof, Saint~Petersburg 198504, Russia}
\date{Accepted 2007 September 25. Received 2007 September 15; in original form 2007 July 14}
\pagerange{\pageref{firstpage}--\pageref{lastpage}}
\pubyear{2007}

\begin{document}

\label{firstpage}

\maketitle

\begin{abstract}
 We show different expressions of distribution functions (DFs)
which depend only on the two classical integrals of the energy and the 
magnitude of the angular momentum with respect to the axis of symmetry 
for stellar systems with known axisymmetric densities. 
The density of the system is required to be a product of 
 functions separable in the potential and the radial coordinate, 
where the functions of the radial coordinate are powers of a sum of a square 
  of the radial coordinate and its unit scale. 
The even part of the two-integral DF corresponding to this type of density  
is in turn a sum or an infinite series of products of functions of the energy and of the 
magnitude of the angular momentum about the axis of symmetry. 
A similar expression of its odd part  
can be also obtained under the assumption of the rotation laws.  
It can be further shown that 
these expressions are in fact equivalent to those 
obtained by using Hunter and Qian's contour integral formulae   
for the system.  
This method is generally computationally preferable to the contour integral method. 
Two examples are given 
to obtain the even and odd parts of their two-integral DFs. 
One is for the prolate Jaffe model  
and the other for the prolate Plummer model. 

It can be also found that the Hunter-Qian contour integral 
formulae of the two-integral even DF for axisymmetric systems can be recovered 
by use of the Laplace-Mellin integral transformation originally developed by Dejonghe.
\end{abstract}

\begin{keywords}
celestial mechanics - stellar dynamics - galaxies.
\end{keywords}

\section{Introduction}
\label{intro}
This paper is a continuity of our recent work (Jiang \& Ossipkov 2007a,b) to  
construct a self-consistent stellar system by means of finding a two-integral 
distribution function (hereafter DF)
 for a stellar system with a known 
gravitational potential.   
Jiang and Ossipkov (2007a) have shown a method of   
finding anisotropic DFs for spherical galaxies. 
This is an combination of Eddington's (1916) formula 
and Fricke's (1952) expansion idea.  Of course, 
they can be also regarded as simply
an extension of the idea of Eddington.  
Some similar formulae are given by Jiang and Ossipkov (2007b) 
for finding two-integral DFs for axisymmetric stellar systems. 

There is a long history of finding the DFs 
for a stellar system with a known gravitational potential. 
Eddington's (1916) formula is for spherical galaxies 
and is in fact a solution to the Abel integral equation 
(see Binney \& Tremaine 1987, hereafter BT;  Dejonghe 1986).
Fricke (1952) gave an expansion idea, that is,   
DFs which are products of the two powers 
of the energy and the square of the angular momentum about the axis of symmetry 
correspond to densities which are proportional to products of the potential 
and the radial coordinate for axisymmetric systems.  
After that, different integral transformation techniques are used to 
obtain the DFs of spherical or axisymmetric stellar systems
(e.g. Lynden-Bell 1962a; Hunter 1975; Kalnajs 1976; Dejonghe 1986). 
Hunter and Qian (1993) gave complex contour integral formulae to get the two-integral DFs 
for axisymmetric systems.  
   
Therefore, in this paper, we present different expressions of the two-integral DF 
in the stellar system with the axisymmetric density that is required to be expressed as  
a product of two functions: one is a function of the potential and the other a 
power of a sum of a square 
  of the radial coordinate and its unit scale.  
We first introduce the fundamental integral equations 
of the problem of finding the two-integral DFs in Section \ref{inteq} and then 
 present our method of solution in Section \ref{2df}. 
We show in Section \ref{even} that 
the Hunter-Qian complex contour integral formulae can be recovered by use of the 
Laplace-Mellin integral transformation defined by Dejonghe (1986) and that the two-integral even DFs for stellar systems
with the known axisymmetric densities of this type mentioned above can be expressed as 
a finite sum or an infinite series of products of functions of the energy and of the 
magnitude of the angular momentum about the axis of symmetry. 
In Section  \ref{odd} we give the similar formulae of the two-integral odd DFs   
under the assumption of the rotation laws. 
In Section \ref{edfpro} we use this device to obtain the two-integral DFs  
for both the prolate Jaffe and Plummer models. These DFs are indeed in accordance with 
those derived from the Hunter-Qian contour integral. 
Section {\ref{con} summarizes our results and exhibits our conclusions. 

\section{The fundamental integral equations}
\label{inteq}
Assume that a stellar system has a gravitational potential with a upper bound. 
Let $\Phi$ and $E$ be, respectively, the potential 
and the energy of a star in the system. 
As in BT, we can choose a  constant $\Phi_0$ 
such that the system has only stars of the energy $E< \Phi_0,$  
and then use the usually so-called relative potential $\psi=-\Phi+\Phi_0$ and energy $\varepsilon=-E+\Phi_0$  
to describe the stellar system. 
Here, $\varepsilon=0$ is a relative energy of escape from the system. 
Given a relative potential $\psi=\psi({\bf r})$ of a stellar system
its mass density $\rho=\rho({\bf r})$ can be derived from Poisson's equation. 
Then the problem of finding the DF $f=f({\bf r},{\bf v})$ of the system is that of solving the following integral equation  
\begin{equation}
 \rho=\int fd^3{\bf v}, \label{int0}
\end{equation}
where ${\bf r}$ is a position vector, ${\bf v}$ is a velocity vector. 
In an axisymmetric system, we generally use 
the cylindrical 
polar coordinates $(R,\varphi,z)$ with  
the $z$-axis being that of symmetry. 
Let ${\bf v}=(v_R,v_\phi,v_z)$ and $L_z$ be, respectively, 
the velocity and the component of angular 
momentum about the $z$-axis.  Then we can know that the relative energy $\varepsilon$ 
and the $z$-axis angular momentum $L_z=R v_\phi$ are two isolating integrals for any orbit in 
the axisymmetric system. Thus, by the Jeans theorem, the DF of a steady-state stellar system 
in an axisymmetric potential can be regarded as a non-negative function of $\varepsilon$ 
and $L_z,$  denoted by $f(\varepsilon,L_z),$ and then 
for an axisymmetric system,  (\ref{int0}) can be rewritten as 
\begin{equation}
\rho=\int  f(\varepsilon,L_z)d^3{\bf v}. \label{inta1}
\end{equation}
Put $v_m=\sqrt{v_R^2+v_z^2}$ and define cylindrical coordinates $(v_m,v_\phi, \theta)$ in velocity space by the following relations:
$v_R=v_m\cos\theta,$ $ v_z=v_m\sin\theta.$ It thus follows that $d^3{\bf v}=v_mdv_mdv_\phi d\theta.$ 
Since $\varepsilon=\psi-(1/2)(v_R^2+v_\phi^2+v_z^2),$  the integrand of 
(\ref{inta1}) is independent of $\theta$  so that  
we integrate out $\theta.$  We then change the variables $(v_m,v_\phi)$ to $(\varepsilon,L_z).$ 
It is easy to see that 
the integral in $dL_z$ is actually an integral of $f$ from $-R\sqrt{2(\psi-\varepsilon)}$ to $R\sqrt{2(\psi-\varepsilon)}.$ 
We finally restrict ourselves    
to only positive values of $L_z$ by changing the lower integration limit 
to $0$ and integrating the function $2f_+(\varepsilon,L_z),$ thus getting
\begin{equation}
 \rho=\frac{4\pi}{R}\int_0^\psi\left[\int_0^{R\sqrt{2(\psi-\varepsilon)}} 
f_+(\varepsilon,L_z)dL_z\right]d\varepsilon \label{inta2}
\end{equation}
for the system having only stars with $\varepsilon\geq0$ [that is, $f(\varepsilon,L_z)=0$ for $\varepsilon<0$], 
where $f_+(\varepsilon,L_z)=[f(\varepsilon,L_z)+f(\varepsilon,-L_z)]/2.$
This implies that a given density determines $f_+(\varepsilon,L_z)$ which is 
just the part of the DF that is even in $L_z.$ 
Hence $f_+(\varepsilon,L_z)$ is usually called  the even DF. 

Using the similar derivation to that of (\ref{inta2}), we can rechange  
$\rho \langle v_\phi\rangle=\int  v_\phi f(\varepsilon,L_z)d^3{\bf v}$ as 
\begin{equation}
 \rho \langle v_\phi\rangle=\frac{4\pi}{R^2}\int_{0}^\psi\left[\int_0^{R\sqrt{2(\psi-\varepsilon)}} 
L_zf_\_(\varepsilon,L_z)dL_z\right]d\varepsilon \label{intaodd}, 
\end{equation}
where $f_\_(\varepsilon,L_z)=[f(\varepsilon,L_z)-f(\varepsilon,-L_z)]/2.$
Usually, $f_\_(\varepsilon,L_z)$ is  called the odd DF of the stellar systems.
Equation (\ref{intaodd}) means that the odd DF can be derived from the rotational velocity. 
We can also find that    
if $\rho$ and $f_+(\varepsilon,L_z)$ in (\ref{inta2}) 
are replaced by $\rho R \langle v_\phi\rangle$ and $L_zf_\_(\varepsilon,L_z),$ respectively, 
then (\ref{inta2}) becomes (\ref{intaodd}) for $L_zf_\_(\varepsilon,L_z).$ 

Similarly, 
put $Q=\varepsilon-L_z^2/(2R_a^2),$ where $R_a$ is a scaling radius, and 
assume that the DF is dependent on $Q$ and $L_z,$ denoted by  $f(Q,L_z),$ 
and that  the system has only stars with $Q\geq 0,$
or equivalently, $f=0$ for $Q<0.$ 
Obviously, $Q\to \varepsilon$ as $R_a \to \infty.$ 
Then, for an axisymmetric system,  the integral equation (\ref{int0}) can be rechanged as 
\begin{equation}
\rho=\int  f(Q,L_z)d^3{\bf v}. \label{intaq1} 
\end{equation}
By changing the variables of the integral in (\ref{intaq1}), it follows that 
\begin{equation}
\rho=\frac{4\pi}{R} \int_0^{\psi}\left[ \int_0^{R\sqrt{2(\psi-Q)/(1+R^2/R_a^2)}}  
f_+(Q,L_z)dL_z\right] dQ, \label{intaq2} 
\end{equation} 
where $f_+(Q,L_z)=[f(Q,L_z)+f(Q,-L_z)]/2.$ 
Naturally, $f_+(Q,L_z)$ is the even part of $f(Q,L_z).$ 
Also, the odd part $f_\_(Q,L_z)=[f(Q,L_z)-f(Q,-L_z)]/2$ of the DF satisfies 
\begin{equation}
 \rho\langle v_\phi\rangle=\frac{4\pi}{R^2}\int_{0}^\psi\left[ \int_0^{R\sqrt{2(\psi-Q)/(1+R^2/R_a^2)}} 
L_zf_\_(Q,L_z)dL_z\right]dQ \label{intaqodd}.
\end{equation}

Obviously, (\ref{intaq1}), (\ref{intaq2}) and (\ref{intaqodd})  coincide with (\ref{inta1}), (\ref{inta2}) and (\ref{intaodd}), respectively.   
In other words, (\ref{inta1}), (\ref{inta2}) and (\ref{intaodd}) are, respectively, 
limits of (\ref{intaq1}), (\ref{intaq2}) and (\ref{intaqodd}) when $R_a\to \infty.$ 
Thus it suffices to consider the case of $f(Q,L_z)$ in the next section. 

Once $f_+(Q,L_z)$ is known, $f(Q,L_z)$ can be obtained 
under some assumptions (e.g. Lynden-Bell 1962a; Miyamoto 1971, 1974;
Ossipkov 1978;  Dejonghe 1986; Kutuzov 1995) 
and further $\rho$ can be easily calculated 
by integration and $\psi$ by solving Poisson's equation for the axisymmetric system. 

The inverse problem that  is now investigated for any axisymmetric system is 
how to derive the two-integral even DF $f_+(Q,L_z)$ from the density $\rho$  
and how to deduce the two-integral odd DF $f_\_(Q,L_z)$ under the assumption of the rotation laws. 

\section{The two-integral DFs}
\label{2df}
In this section, we shall show different expressions of the even and odd parts of the two-integral DFs   
corresponding to three types of the mass densities given by 
\begin{equation}
\rho(\psi,R)=\hat{\rho}(\psi)/(1+R^2/R_a^2),
\label{rho1}
\end{equation}  
\begin{equation}
\rho(\psi,R)=\tilde{\rho}(\psi)/(1+R^2/R_a^2)^{1/2},
\label{rho1t}
\end{equation}  
\begin{equation}
\rho(\psi,R)=\bar{\rho}(\psi)/(1+R^2/R_a^2)^{3/2}.
\label{rho1b}
\end{equation}  
We shall also give the similar expressions of the even and odd DFs for the spherical mass density 
$\rho=\check{\rho}(\psi).$

\subsection{The even DFs}
\label{even}
We can know that under some suitable assumptions 
a complex integral expression of the two-integral DFs corresponding to the above three types 
of the axisymmetric mass densities can be obtained by use of 
Hunter and Qian's (1993) complex contour formula. 
It can be further known that this contour integral is the same as that derived from 
Dejonghe's (1986) Laplace-Mellin integral transformation for the axisymmetric mass density of the 
separable form $\rho(\psi,R)=F(\psi)g(R),$ where $F(\psi)$ and $g(R)$ are two functions such that 
both the Laplace and Mellin integral transformations of $F(\psi)$ and $g(R)$ are valid, respectively.  
To show this, we first have to recall the following definitions of the Laplace-Mellin integral transformations
\begin{equation}
{\mathfrak L}_{\varepsilon\to\alpha}{\mathfrak M}_{L_z\to\beta}\{f_+\}
=\int\limits_{0}^{+\infty}e^{-\alpha\varepsilon}d\varepsilon
\int\limits_{0}^{+\infty}L_z^{\beta-1}f_+(\varepsilon, L_z)dL_z
\label{dflm}
\end{equation} 
and 
\begin{equation}
{\mathfrak L}_{\psi\to\alpha}{\mathfrak M}_{R\to\beta}\{\rho\}
=\int\limits_{0}^{+\infty}e^{-\alpha\psi}d\psi
\int\limits_{0}^{+\infty}R^{\beta-1}\rho(\psi, R)dR
\label{rholm}
\end{equation} 
with their inversions
\begin{equation}
f_+(\varepsilon, L_z)=-\frac{1}{4\pi^2}
\int\limits_{\alpha_0-{\rm i}\infty}^{\alpha_0+{\rm i}\infty}e^{\alpha\varepsilon}d\alpha
\int\limits_{\beta_0-{\rm i}\infty}^{\beta_0+{\rm i}\infty}L_z^{-\beta}
{\mathfrak L}_{\varepsilon\to\alpha}{\mathfrak M}_{L_z\to\beta}\{f_+\}d\beta
\label{dflminverse}
\end{equation} 
and 
\begin{equation}
\rho(\psi,R)=-\frac{1}{4\pi^2}
\int\limits_{\alpha_0-{\rm i}\infty}^{\alpha_0+{\rm i}\infty}e^{\alpha\psi}d\alpha
\int\limits_{\beta_0-{\rm i}\infty}^{\beta_0+{\rm i}\infty}R^{-\beta}
{\mathfrak L}_{\psi\to\alpha}{\mathfrak M}_{R\to\beta}\{\rho\}d\beta,
\label{rholminverse}
\end{equation}
where, $\alpha_0$ and $\beta_0$ are two suitable real numbers, 
${\mathfrak L}$ and ${\mathfrak M}$ denote the Laplace and Mellin integral transformations, respectively. 
Here and below, ${\rm i}$ is an imaginary unit.  
Obviously,  the Laplace-Mellin integral transformations of (\ref{rho1}), (\ref{rho1t}) and (\ref{rho1b}) are valid for ${\rm Re}(\beta)<1$ 
when $\hat{\rho}(\psi),$ $\tilde{\rho}(\psi)$ and $\bar{\rho}(\psi)$ are suitable for the Laplace transformation.

We can then know that the Laplace-Mellin integral transformations of 
the mass density $\rho=\rho(\psi,R)$ and its DF $f_+=f_+(\varepsilon, L_z)$ have the 
following relation [see Dejonghe 1986, Page 273, (2.3.3)]
\begin{equation}
{\mathfrak L}_{\varepsilon\to\alpha}{\mathfrak M}_{L_z\to\beta}\{f_+\}=
\frac{\alpha^{(3-\beta)/2}2^{\beta/2}}{2^{3/2}\pi\Gamma\left(\frac{1-\beta}{2}\right)}
{\mathfrak L}_{\psi\to\alpha}{\mathfrak M}_{R\to\beta}\{\rho\},
\label{rel-df-rho}
\end{equation}
where $\Gamma(x)$ denotes the Gamma function, $-\pi/2<\arg(\alpha)<\pi/2$ and ${\rm Re}(\beta)<1.$ 
Note that the Laplace-Mellin integral transformation of the separable density $\rho(\psi,R)=F(\psi)g(R)$ is given by 
${\mathfrak L}_{\psi\to\alpha}{\mathfrak M}_{R\to\beta}\{\rho\}={\mathfrak L}_{\psi\to\alpha}\{F\}{\mathfrak M}_{R\to\beta}\{g\}.$ 
Thus the Mellin integral transformation of $f_+=f_+(\varepsilon, L_z)$ can be obtained as
\begin{equation}
{\mathfrak M}_{L_z\to\beta}\{f_+\}=
\frac{{\mathfrak M}_{R\to\beta}\{g\}2^{\beta/2}}{2^{3/2}\pi\Gamma\left(\frac{1-\beta}{2}\right)}
{\mathfrak L}_{\alpha\to\varepsilon}^{-1}\left\{\alpha^{(3-\beta)/2}{\mathfrak L}_{\psi\to\alpha}\{F\}\right\},
\label{df-mellin}
\end{equation}
where ${\mathfrak L}_{\alpha\to\varepsilon}^{-1}$ denotes the inversion of the Laplace integral transformation. 
It can be shown that the Laplace inversion transformation 
on the right side of (\ref{df-mellin}) has the following form  
(see Dejonghe 1986, Page 306, Appendix A1, Theorem 1)
\begin{equation}
{\mathfrak L}_{\alpha\to\varepsilon}^{-1}\left\{\alpha^{(3-\beta)/2}{\mathfrak L}_{\psi\to\alpha}\{F\}\right\}
=\frac{1}{\Gamma\left(\frac{1+\beta}{2}\right)}
\frac{d^2}{d \varepsilon^2}\int_0^\varepsilon F(\psi)(\varepsilon-\psi)^{(\beta-1)/2}d\psi,
\label{laplace-inverse}
\end{equation}
where $1>{\rm Re}(\beta)>-1.$ Assume that $F(0)=F^{\prime}(0)=0.$ Then (\ref{laplace-inverse}) gives
\begin{equation}
{\mathfrak L}_{\alpha\to\varepsilon}^{-1}\left\{\alpha^{(3-\beta)/2}{\mathfrak L}_{\psi\to\alpha}\{F\}\right\}
=\frac{1}{\Gamma\left(\frac{1+\beta}{2}\right)}
\int_0^\varepsilon \frac{d^2 F}{d \psi^2}(\psi)(\varepsilon-\psi)^{(\beta-1)/2}d\psi.
\label{laplace-inverse2}
\end{equation}
Assume without loss of generality that $F(\psi)$ has an analytic continuation near the real $\psi$-axis in the complex $\psi$-plane. 
Then equation (\ref{laplace-inverse2}) can be rewritten as
\begin{equation}
{\mathfrak L}_{\alpha\to\varepsilon}^{-1}\left\{\alpha^{(3-\beta)/2}{\mathfrak L}_{\psi\to\alpha}\{F\}\right\}
=-\frac{e^{- {\rm i}\frac{\beta\pi}{2}}}{2\cos\left(\frac{\beta\pi}{2}\right)\Gamma\left(\frac{1+\beta}{2}\right)}
\int_{C(\varepsilon)}\frac{d^2 F}{d \psi^2}(\psi)(\varepsilon-\psi)^{(\beta-1)/2}d\psi, 
\label{laplace-inverse3}
\end{equation}
where and below,  $C(\varepsilon)$ is a suitable contour 
of the complex integral (\ref{laplace-inverse3}) with respect to the 
complex variable $\psi;$ the contour is a loop which starts 
from the lower side of the real $\psi$-axis at $\psi=0,$ 
passing through a physically considered window ${\cal P}$ in the real $\psi$-axis to the right of $\psi=\varepsilon,$ 
to the upper side of the real $\psi$-axis at $\psi=0;$ 
the window ${\cal P}$ physically considered must lie between the relative energy $\varepsilon$ and the maximal relative potential $\psi_0.$
Figure~\ref{contour} shows one cut (broken line), one window ${\cal P}$ and one contour $C(\varepsilon).$ 
\begin{figure}
\centering 
\psfrag{0}[rb]{{$ 0$}}
\psfrag{C}{{$ C(\varepsilon)$}}
\psfrag{E}{{$ \varepsilon$}}
\psfrag{Im(psi)}{{$ {\rm Im}(\psi)$}}
\psfrag{psi0}{{$ \psi_0$}}
\psfrag{psi}{{$ {\rm Re}(\psi)$}}
\psfrag{F}{{${\cal P}$}}
\includegraphics[bb=0 0 707 471,totalheight=50mm,width=0.5\linewidth]{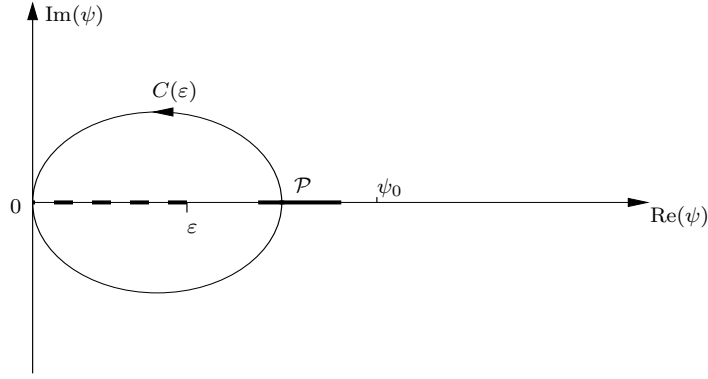}
\caption{The contour $C(\varepsilon)$ of the complex integral given by (\ref{laplace-inverse3}). Both the   
branch point and the pole of its integrand are the relative energy $\varepsilon$ in the real $\psi$-axis to the left side of the 
maximal relative potential $\psi_0$.}
\label{contour}
\end{figure} 
Inserting (\ref{laplace-inverse3}) into (\ref{df-mellin}), we can get 
\begin{equation}
{\mathfrak M}_{L_z\to\beta}\{f_+\}=
-\frac{1}{4\sqrt{2}\pi}
\int_{C(\varepsilon)}\frac{d^2 F}{d \psi^2}(\psi)\frac{1}{\sqrt{\varepsilon-\psi}}
\frac{{\mathfrak M}_{R\to\beta}\{g\}[2(\varepsilon-\psi)]^{\beta/2}e^{- {\rm i}\frac{\beta\pi}{2}}}
{\Gamma\left(\frac{1-\beta}{2}\right)\Gamma\left(\frac{1+\beta}{2}\right)\cos\left(\frac{\beta\pi}{2}\right)}
d\psi.
\label{df-mellin2}
\end{equation}
By using the identity equality that 
$\Gamma\left(\frac{1-\beta}{2}\right)\Gamma\left(\frac{1+\beta}{2}\right)\cos\left(\frac{\beta\pi}{2}\right)=\pi$ 
(see Gradshteyn \& Ryzhik 1965, Page 937, 8.334, 2),
the Mellin inverse integral transformation of equation (\ref{df-mellin2}) can be expressed as 
\begin{equation}
f_+(\varepsilon, L_z)=
-\frac{1}{4\sqrt{2}\pi^2}
\int_{C(\varepsilon)}\frac{d^2 F}{d \psi^2}(\psi)\frac{1}{\sqrt{\varepsilon-\psi}}
{\mathfrak M}_{\beta\to L_z}^{-1}\left\{{\mathfrak M}_{R\to\beta}\{g\}[2(\varepsilon-\psi)]^{\beta/2}e^{- {\rm i}\frac{\beta\pi}{2}}\right\}
d\psi 
\label{df-mellin-inverse}
\end{equation}
where ${\mathfrak M}_{\beta\to L_z}^{-1}$ denotes the inversion of the Mellin integral transformation. 
(\ref{df-mellin-inverse}) can be equivalently rewritten as
\begin{eqnarray}
f_+(\varepsilon, L_z)\mathop{=}\limits^{({\rm a})}-\frac{1}{4\sqrt{2}\pi^2}
\int_{C(\varepsilon)}\frac{d^2 F}{d \psi^2}(\psi)\frac{1}{\sqrt{\varepsilon-\psi}}
{\mathfrak M}_{\beta\to \sqrt{L_z^2/[2(\varepsilon-\psi)]}}^{-1}
\left\{{\mathfrak M}_{R\to\beta}\{g\}e^{- {\rm i}\frac{\beta\pi}{2}}\right\}
d\psi \nonumber \\
\mathop{=}\limits^{({\rm b})}-\frac{1}{4\sqrt{2}\pi^2}
\int_{C(\varepsilon)}\frac{d^2 F}{d \psi^2}(\psi)\frac{1}{\sqrt{\varepsilon-\psi}}
g\left( {\rm i}\sqrt{\frac{L_z^2}{2(\varepsilon-\psi)}}\right)
d\psi 
\label{df-mellin2inverse}
\end{eqnarray}
where (a) results from the definition of the Mellin integral transformation and (b) is 
obtained by virtual of  the fact that 
${\mathfrak M}_{r\to p}\left\{g(re^{{\rm i}\theta})\right\}=e^{-{\rm i}p\theta}{\mathfrak M}_{r\to p}\left\{g(r)\right\}$
[see Davies 2002, Page 203, (12.12)]. 
Hence (\ref{df-mellin2inverse}) is in accordance with the Hunter-Qian complex contour integral (Hunter \& Qian 1993)
when the mass density $\rho(\psi,R)=F(\psi)g(R)$ is any one of the three densities 
defined by (\ref{rho1}), (\ref{rho1t}) and (\ref{rho1b}). It can be similarly shown that 
the two-integral even DF
\begin{eqnarray}
f_+(\varepsilon, L_z)=-\frac{1}{4\sqrt{2}\pi^2}
\int_{C(\varepsilon)}\frac{\partial^2 \rho}{\partial \psi^2}
\left(\psi, {\rm i}\sqrt{\frac{L_z^2}{2(\varepsilon-\psi)}}\right)\frac{d\psi}{\sqrt{\varepsilon-\psi}}
\label{df-mellin3inverse}
\end{eqnarray}
corresponds to the general axisymmetric mass density $\rho=\rho(\psi,R)$   
having an analytic continuation in the complex plane with respect to the variable $\psi$ and satisfying that 
$\rho(0,R)
=[\partial  \rho(\psi,R)/{\partial \psi}]_{\psi=0}=0.$ Obviously, 
(\ref{df-mellin3inverse}) is the same as the complex contour integral formulae mentioned above.  

Let $\mu$ and $\nu$ be two constants. 
Assume that $-\mu-\nu$ is not a natural number and that $p\geq 3/2.$  
If the two functions $F$ and $g$ in (\ref{df-mellin}) are defined as 
$F(\psi)=\psi^p$ and $g(R)=(R^2/R_a^2)^\mu/(1+R^2/R_a^2)^{\mu+\nu},$ respectively, 
then ${\mathfrak L}_{\psi\to\alpha}\{F\}=\Gamma(p+1)/\alpha^{p+1}$ 
and ${\mathfrak M}_{R\to\beta}\{g\}=(R_a^2)^{\beta/2}\Gamma(\mu+\beta/2)\Gamma(\nu-\beta/2)/[2\Gamma(\mu+\nu)].$ 
Thus (\ref{df-mellin}) becomes 
\begin{eqnarray}
{\mathfrak M}_{L_z\to\beta}\{f_+\}=
\frac{\Gamma(p+1)}{2^{5/2}\pi\Gamma(\mu+\nu)}
\frac{\Gamma\left(\mu+\frac{\beta}{2}\right)\Gamma\left(\nu-\frac{\beta}{2}\right)(2R_a^2)^{\beta/2}}{\Gamma\left(\frac{1-\beta}{2}\right)}
{\mathfrak L}_{\alpha\to\varepsilon}^{-1}\left\{\alpha^{(1-\beta)/2-p}\right\} \nonumber \\
=\frac{\Gamma(p+1)\varepsilon^{p-3/2}}{2^{5/2}\pi\Gamma(\mu+\nu)}
\frac{\Gamma\left(\mu+\frac{\beta}{2}\right)\Gamma\left(\nu-\frac{\beta}{2}\right)(2\varepsilon R_a^2)^{\beta/2}}
{\Gamma\left(\frac{1-\beta}{2}\right)\Gamma\left(p-\frac{1}{2}+\frac{\beta}{2}\right)}.
\label{df-mellin-power}
\end{eqnarray}
Then (\ref{df-mellin-power}) can be equivalently rewritten as 
\begin{eqnarray}
f_+(\varepsilon,L_z)=\frac{\Gamma(p+1)\varepsilon^{p-3/2}}{2^{5/2}\pi\Gamma(\mu+\nu)}
\frac{1}{2\pi {\rm i}} \int\limits_{\beta_0-{\rm i}\infty}^{\beta_0+{\rm i}\infty}
\frac{\Gamma\left(\mu+\frac{\beta}{2}\right)\Gamma\left(\nu-\frac{\beta}{2}\right)}
{\Gamma\left(\frac{1-\beta}{2}\right)\Gamma\left(p-\frac{1}{2}+\frac{\beta}{2}\right)}
\left(\frac{L_z^2}{2\varepsilon R_a^2}\right)^{-\beta/2}d\beta \nonumber \\
=\frac{\Gamma(p+1)\varepsilon^{p-3/2}}{2^{3/2}\pi\Gamma(\mu+\nu)}
{\mathbb H}\left(\mu,\nu,p-\frac{1}{2},\frac{1}{2};\frac{L_z^2}{2\varepsilon R_a^2}\right),
\label{df-mellin-power2}
\end{eqnarray}
where ${\mathbb H}(\mu,\nu,c,d;x)$ is defined by 
\begin{equation}
{\mathbb H}(\mu,\nu,\xi,\eta;x)=\frac{1}{2\pi{\rm i}}\int_{C}
\frac{\Gamma(\mu+s)\Gamma(\nu-s)}{\Gamma(\xi+s)\Gamma(\eta-s)}x^{-s}ds
\label{hfun}
\end{equation} 
with a contour $C$ determined by $\beta_0$ such that $-\mu$ are on its left side 
and $\nu$ on its right side. 
In the case that $\mu+\eta$ and $\nu+\xi$ are not negative integers, the complex integral (\ref{hfun}) can be calculated 
and expressed as follows. When $0\leq x<1,$ if $\mu-\xi$ is a nonnegative integer, then ${\mathbb H}(\mu,\nu,\xi,\eta;x)=0,$  or else 
${\mathbb H}(\mu,\nu,\xi,\eta;x)=x^\mu  \hbox{}_2F_1(\mu+\nu,1+\mu-\xi;\mu+\eta;x)\Gamma(\mu+\nu)/[\Gamma(\xi-\mu)\Gamma(\mu+\eta)];$ 
when $x>1,$ if $\nu-\eta$ is a nonnegative integer, then ${\mathbb H}(\mu,\nu,\xi,\eta;x)=0,$  or else
${\mathbb H}(\mu,\nu,\xi,\eta;x)=x^{-\nu}  \hbox{}_2F_1(\mu+\nu,1+\nu-\eta;\nu+\xi;1/x)\Gamma(\mu+\nu)/[\Gamma(\eta-\nu)\Gamma(\nu+\xi)].$ 
Here, $\hbox{}_2F_1$ is a hypergeometric function. 
In particular, ${\mathbb H}(0,1,p-1/2,1/2;x)=\hbox{}_2F_1(1,3/2-p;1/2;x)/[\Gamma(p-1/2)\Gamma(1/2)]$ for any $x\in [0,1),$ 
and ${\mathbb H}(0,1,p-1/2,1/2;x)=x^{-1}\hbox{}_2F_1(1,3/2;p+1/2;1/x)/[\Gamma(p+1/2)\Gamma(-1/2)]$ for any $x\in (1,+\infty).$ 

We shall below give different expressions of the two-integral DFs for the systems.    
First, let us consider the two-integral even DFs corresponding to the mass density defined by (\ref{rho1}). 
Assume that $\hat{\rho}(\psi)$ in (\ref{rho1}) can be expressed as a convergent power series of the form 
$\hat{\rho}(\psi)=\sum_{n=p}^{+\infty}\hat{a}_n\psi^n$  
for $p\geq 3/2.$ Then the mass density defined by (\ref{rho1}) can be rewritten as 
\begin{equation}
\rho(\psi,R)=\sum\limits_{n=p}^{+\infty}\hat{a}_n\psi^n/(1+R^2/R_a^2). 
\label{rho1h2}
\end{equation} 
By (\ref{df-mellin-power2}), it can be found that 
the even DF corresponding to the mass density (\ref{rho1h2}) is 
\begin{eqnarray}
f_+(\varepsilon, L_z)=\sum\limits_{n=p}^{+\infty}
\frac{\hat{a}_n\Gamma(n+1)\varepsilon^{n-3/2}}{2^{3/2}\pi}
{\mathbb H}\left(0,1,n-\frac{1}{2},\frac{1}{2};\frac{L_z^2}{2\varepsilon R_a^2}\right) \hspace*{5cm}\nonumber \\
=\left\{\begin{array}{cc}
\sum\limits_{n=p}^{+\infty}
\frac{\hat{a}_n\Gamma(n+1)\varepsilon^{n-3/2}}{(2\pi)^{3/2}\Gamma(n-1/2)}
\hbox{ }_2F_1\left(1,\frac{3}{2}-n;\frac{1}{2};\frac{L_z^2}{2\varepsilon R_a^2}\right) & {\rm as~~~} \varepsilon\geq L_z^2/(2R_a^2),  \\
-\sum\limits_{n=p}^{+\infty}
\frac{\hat{a}_n\Gamma(n+1)\varepsilon^{n-3/2}}{2(2\pi)^{3/2}\Gamma(n+1/2)}
\left(\frac{L_z^2}{2\varepsilon R_a^2}\right)^{-1}
\hbox{}_2F_1\left(1,\frac{3}{2};n+\frac{1}{2};\frac{2\varepsilon R_a^2}{L_z^2}\right) & {\rm as~~~} \varepsilon\leq L_z^2/(2R_a^2).
\end{array}\right.
\label{dfaq2}
\end{eqnarray} 

It is worth mentioning that the series in (\ref{dfaq2}) 
are convergent to the corresponding contour integrals 
obtained by virtual of the Hunter-Qian contour integral formulae for the systems since 
all the terms in this series can be equivalently rewritten 
as the corresponding Hunter-Qian contour integrals. 
If the convergence of the above power series is fast enough, then one can 
adopt a sum of finite previous terms of the series (\ref{dfaq2}) as a viable approximation of the even DF.  
Note that it is easy to calculate numerically these hypergeometric functions in (\ref{dfaq2}).  
Hence this method is in general computationally preferable to the contour integral method 

By using the method given by Jiang and Ossipkov (2007b), 
it can be easily known that  the stellar system with the mass density of the form (\ref{rho1t}) 
has  the two-integral even DF $f_+(Q)$ of the form  
\begin{equation}
f_+(Q)=
\frac{1}{\sqrt{8}\pi^2}\frac{d^{2}}{dQ^{2}}\int_0^Q
\frac{\tilde{\rho}(\psi)d\psi}{\sqrt{Q-\psi}}
\label{dfaqt}
\end{equation} 
since $\tilde{\rho}(\psi)$ vanishes at $\psi=0$ for the stellar system. 

Then we shall show the two-integral even DF corresponding to the mass density (\ref{rho1b}).  
Note that (\ref{rho1b}) can be rechanged as 
\begin{equation}
\rho(\psi,R)=\bar{\rho}(\psi)[R_a^2/(1+R^2/R_a^2)^{1/2}-R^2/(1+R^2/R_a^2)^{3/2}]/R_a^2.  
\label{rho1b2}
\end{equation}
Assume that $[d^j\bar{\rho}(\psi)/d\psi^j]_{\psi=0}=0$ for $j=0,1.$ 
Similarly, it then follows that  the two-integral even DF
\begin{equation}
f_+(Q, L_z)=
\frac{1}{\sqrt{8}\pi^2}\frac{d^{2}}{dQ^{2}}\int_0^Q
\frac{\bar{\rho}(\psi)d\psi}{\sqrt{Q-\psi}}
-\frac{L_z^2}{\sqrt{8}\pi^2R_a^2}\frac{d^{3}}{dQ^{3}}\int_0^Q
\frac{\bar{\rho}(\psi)d\psi}{\sqrt{Q-\psi}}
\label{dfaqb}
\end{equation} 
corresponds to the mass density (\ref{rho1b2}) (or say, (\ref{rho1b})).  

It can be also proven that (\ref{dfaqt}) and (\ref{dfaqb}) are in accordance with those 
obtained by use of Hunter and Qian's (1993) contour integral formula 
and that they can be derived using 
Dejonghe's (1986) Laplace-Mellin integral transformation. 
Obviously, these real integrals are easier to calculate 
than the corresponding complex contour integrals. 

We can finally mention that for the spherical stellar system with the mass density $\rho=\check{\rho}(\psi),$   
 there is a unique isotropic DF of the Eddington (1916) type 
\begin{equation}
f_+(\varepsilon)=
\frac{1}{\sqrt{8}\pi^2}\frac{d^{2}}{d\varepsilon^{2}}\int_0^\varepsilon
\frac{\check{\rho}(\psi)d\psi}{\sqrt{\varepsilon-\psi}}
\label{dfs}
\end{equation}
and  an infinity of different anisotropic DFs of the Ossipkov-Merritt type  
 corresponding to an infinity of  various values of an anisotropy radius (Ossipkov 1979; Merritt 1985). 

\subsection{The odd DFs}
\label{odd}
Note that if $\rho$ and $f_+(Q,L_z)$ in (\ref{intaq2}) 
are replaced by $\rho R \langle v_\phi\rangle$ and $L_zf_\_(Q,L_z),$ respectively, 
then (\ref{intaq2}) becomes (\ref{intaqodd}) for $L_zf_\_(Q,L_z).$ 
Then we can use Hunter and Qian's contour integral 
(or Dejonghe's Laplace-Mellin transformation, mentioned above) to get 
the odd part of the two-integral DFs corresponding to the three types of the axisymmetric 
mass densities. We shall below show other expressions of the two-integral odd DFs.  

Assume that the rotational velocity of the stellar system obeys 
the rotation law: $\langle{v}_\phi\rangle=R^{2k}/(1+R^2/R_a^2)^k,$ 
where $k$ is a positive constant.  
In the case of $k$ being a natural number, 
this law has been used to get the odd DFs of Binney's (BT) logarithmic model (Jiang \& Ossipkov 2007b). 
When $k=1,$ it is one of the rotation laws considered by Evans (1993). 
Define $[k]=k-k_0,$ where $[k]$ represents a nonnegative integer and $k_0\in [0,1).$ 
Suppose that $\hat{\rho}(\psi)$ in (\ref{rho1}) 
satisfies that $[d^j\hat{\rho}(\psi)/d\psi^j]_{\psi=0}=0$ for $j=0,1,\cdots, [k]+1$ and that the system has only stars with $Q>0.$ 
Take $f_\_(Q,L_z)={\rm sgn}(L_z)L_z^{2k}h(Q)$ as a two-integral odd DF for the mass density defined by (\ref{rho1}). 
Then inserting (\ref{rho1}) into 
the integral equation (\ref{intaqodd}) gives  
\begin{equation}
\frac{R^{2k}\hat{\rho}(\psi)}{(1+R^2/R_a^2)^{k+1}}
=\frac{4\pi 2^{k}R^{2k}}{(k+1)(1+R^2/R_a^2)^{k+1}}
\int_0^\psi h(Q) (\psi-Q)^{k+1}dQ. \label{intaqodd2}
\end{equation} 
It can thus be found from (\ref{intaqodd2}) that 
\begin{equation}
\hat{\rho}(\psi)
=\frac{4\pi 2^{k}}{(k+1)}
\int_0^\psi h(Q) (\psi-Q)^{k+1}dQ. \label{intaqodd3}
\end{equation} 
Using the Laplace integral transformation (Dejonghe 1986), we can get the following inverse formula of 
(\ref{intaqodd3}):  
 \begin{equation}
 h(Q)=\frac{k+1}{4\pi 2^k\Gamma(k+2)\Gamma(1-k_0)}\frac{d^{[k]+3}}{dQ^{[k]+3}}\int_0^Q 
\frac{\hat{\rho}(\psi)d\psi}{(Q-\psi)^{k_0}}.
 \label{solutionaqodd}
\end{equation}
Therefore we can know from (\ref{solutionaqodd}) that 
under the above assumptions the two-integral odd DF is given by 
 \begin{equation}
f_\_(Q,L_z)=\frac{(k+1){\rm sgn}(L_z)L_z^{2k}}{4\pi 2^k\Gamma(k+2)\Gamma(1-k_0)}\frac{d^{[k]+3}}{dQ^{[k]+3}}\int_0^Q 
\frac{\hat{\rho}(\psi)d\psi}{(Q-\psi)^{k_0}}
 \label{dfaqodd}
\end{equation}
for the mass density (\ref{rho1}). 
In particular, when $k$ is a natural number, the two-integral odd DF (\ref{dfaqodd}) for the mass density (\ref{rho1}) can be rewritten as 
 \begin{equation}
f_\_(Q,L_z)=\frac{{\rm sgn}(L_z)L_z^{2k}}{4\pi 2^kk!}\frac{d^{k+2}\hat{\rho}(Q)}{dQ^{k+2}} 
 \label{dfaqodd0}
\end{equation}
on the condition that  $[d^j\hat{\rho}(\psi)/d\psi^j]_{\psi=0}=0$ for $j=0,1,\cdots, k+2.$

Next, we consider another expression of   
the two-integral odd DF corresponding to the mass density of the form 
\begin{equation}
\rho(\psi,R)=\sum\limits_{n=p}^{+\infty}
a_n\psi^n/(1+R^2/R_a^2)^{\gamma}.
\label{rhoall}
\end{equation} 
for any given nonnegative constant $\gamma.$ 
Note that if $\rho$ and $f_+(\varepsilon,L_z)$ in (\ref{inta2}) 
are replaced by $\rho R \langle v_\phi\rangle$ and $L_zf_\_(\varepsilon,L_z),$ respectively, 
then (\ref{inta2}) becomes (\ref{intaodd}) for $L_zf_\_(\varepsilon,L_z).$ 
Hence, by (\ref{df-mellin-power2}), we can find that 
the two-integral odd DF for the mass density (\ref{rhoall}) is 
 \begin{eqnarray}
f_\_(\varepsilon,L_z)=\sum\limits_{n=p}^{+\infty}
\frac{a_n R_a^{2k+1}\Gamma(n+1)\varepsilon^{n-3/2}{\rm sgn}(L_z)|L_z|^{-1}}
{2^{3/2}\pi\Gamma(k+\gamma)}
{\mathbb H}\left(k+\frac{1}{2},\gamma-\frac{1}{2},n-\frac{1}{2},\frac{1}{2};\frac{L_z^2}{2\varepsilon R_a^2}\right) 
\nonumber \\ 
=\frac{R_a^{2k}{\rm sgn}(L_z)}{4\pi \Gamma(k+\gamma)}
\sum\limits_{n=p}^{+\infty}
a_n n! \varepsilon^{n-2}
{\mathbb H}\left(k,\gamma,n-1,1;\frac{L_z^2}{2\varepsilon R_a^2}\right) 
 \label{dfaqoddall}
\end{eqnarray} 
under the above assumption of the rotation law.  The above last equality is obtained by using the following identity equality: 
\begin{eqnarray}
\varepsilon^{n-3/2}|L_z|^{-1}
{\mathbb H}\left(\mu,\nu,\xi,\eta;\frac{L_z^2}{2\varepsilon R_a^2}\right)
=\varepsilon^{n-2}(2R_a^2)^{-1/2}
{\mathbb H}\left(\mu-\frac{1}{2},\nu+\frac{1}{2},\xi-\frac{1}{2},\eta+\frac{1}{2};\frac{L_z^2}{2\varepsilon R_a^2}\right). 
\label{ie}\nonumber
\end{eqnarray}
In (\ref{dfaqoddall}), 
 the functions ${\mathbb H}$ can be expressed in terms of 
hypergeometric functions and so they are easy to calculate numerically. 
This expression (\ref{dfaqoddall}) is suitable for the mass densities defined by (\ref{rho1})-(\ref{rho1b}), 
even for the spherical mass density $\rho=\check{\rho}(\psi).$ 

It can be finally found that all the above series expressions of the odd DF are also convergent to 
the corresponding complex contour integrals obtained by using 
Hunter and Qian's (1993) contour integral formulae. 
If the convergence of the series (\ref{rhoall}) is fast enough, then we can use a sum of 
finite previous terms of (\ref{dfaqoddall}) as a good approximation of the odd DF. 

\section{Application to the prolate models}
\label{edfpro}
In this section, we shall apply the above series expressions into 
getting the even and odd parts of the DFs for both the prolate Jaffe and Plummer models. 

\subsection{The prolate Jaffe model}
\label{edfjaffe}
The gravitational potential of the prolate Jaffe model is as follows: 
\begin{equation}
\Phi(R,z)=\frac{GM}{r_J}
\ln\left(\frac{\sqrt{(\sqrt{R^2+a^2}+b)^2+z^2}}
{\sqrt{(\sqrt{R^2+a^2}+b)^2+z^2}+r_J}\right).
\label{Phipj}
\end{equation}
Here, and everywhere below, $a,$ $b,$ $M$ and $r_J$ are positive constants and
 $G$ is the gravitational constant. 
By  using Poisson's 
equation $\bigtriangledown^2 \Phi=4\pi G\rho,$ it is easy to show from 
(\ref{Phipj}) that 
\begin{equation}
\rho(R,z)=\frac{M}{4\pi r_J}\frac{%\left\{\begin{array}{c}
r_J^2\tau^2(X^3+bX^2+a^2b)+
br_J\tau^3(X^2+a^2) %\\ 
+(3\tau+2r_J)a^2r_JX(X+b)^2 %\end{array}\right\}
}
{\tau^4(\tau+r_J)^2X^3},
\label{rhopj}
\end{equation}
where and below, $\tau=\sqrt{(X+b)^2+z^2}$ and $X=\sqrt{R^2+a^2}.$  
This density-potential pair was given by Jiang and Moss (2002) 
and a two-integral even DF corresponding to this pair was shown by calculating
the Hunter-Qian contour integral. We can below express  
the two-integral even DF as a series of hypergeometric functions, and also obtain an infinity 
of the two-integral odd DFs for this pair under the assumption of the rotation laws. 
It is worthwhile to mention that such equipotential extensions have been studied 
for the Jaffe (1983) spherical model 
(e.g. Jiang 2000; Jiang, Fang, Liu, Moss 2002; Jiang, Fang, Moss 2002) 
and for the spherical $\gamma$ model (Jiang \& Ossipkov 2006). 
The $\gamma$ model was found independently by Kuzmin et al.~(1986), Dehnen (1993), Saha (1993) 
and Tremaine et al.~(1994). It is in fact an extension of the Jaffe model. 
The device of extending from the spherical models to the axisymmetric models 
was originally developed by Miyamoto and Nagai (1975) 
and  independently by Kutuzov and Ossipkov (1976). 
Other similar extensions can be also found in their further work (Nagai \& Miyamoto 1976; 
Kutuzov 1989; Ossipkov 1997; Ossipkov \& Jiang 2007a,b). 

Put $\psi=-\Phi(R,z).$  
Then, by (\ref{Phipj}), the mass density (\ref{rhopj}) can be expressed 
as a function $\rho(\psi,R)$ of $\psi$ and $R,$ that is, it is of the following form
\begin{equation}
\rho(\psi,R)=\check{\rho}(\psi)
+\tilde{\rho}(\psi)/X 
+\hat{\rho}(\psi)/X^2
+\bar{\rho}(\psi)/X^3,
\label{rhopj2}
\end{equation}
where and below,  
$\check{\rho}(\psi),$ $\tilde{\rho}(\psi),$ $\hat{\rho}(\psi)$ and $\bar{\rho}(\psi)$ are denoted by 
\begin{equation}
\check{\rho}(\psi)=\frac{M}{4\pi r_J^5}\frac{
r_J^2w^4+a^2(3+2w)w^5}{(w+1)^2},
\label{rhopjs}
\end{equation}
\begin{equation}
\tilde{\rho}(\psi)=\frac{M}{4\pi r_J^5}\frac{r_J^2bw^3(w+1)
+2a^2b(3+2w)w^5}{(w+1)^2},
\hat{\rho}(\psi)=\frac{M}{4\pi r_J^5}\frac{a^2b^2(3+2w)w^5}
{(w+1)^2}, 
\bar{\rho}(\psi)=\frac{M}{4\pi r_J^5}\frac{r_J^2a^2bw^3}
{(w+1)}
\label{rhopj2htb}
\end{equation}
with $w=e^{r_J\psi/(GM)}-1.$ 
On the right side of (\ref{rhopj2}), the first term is ``spherical'' and the other ones are axisymmetric 
like three types of the mass densities defined by (\ref{rho1}), (\ref{rho1t}) and (\ref{rho1b}). 
The prolate Jaffe model can be hence regarded as a stellar system 
composed of a ``spherical'' component and three axisymmetric ones.  
Thus the DFs for the ``spherical'' and the three axisymmetric subsystems 
can be derived by using the formulae given in the previous sections and so  
the DF of the prolate Jaffe model can be expressed as a sum of the 
DFs of these subsystems. 

Throughout this paper, let $H(x)$ denote the Heaviside step function, that is, 
$H(x)= 1$ as $x\geq 0$ and $H(x)=0$ as $x<0.$  
First, the two-integral even DF of the ``spherical'' subsystem can be expressed as 
\begin{equation}
\check{f}_+(\varepsilon)=
\frac{H(\varepsilon)}{\sqrt{8}\pi^2}\frac{d^{2}}{d\varepsilon^{2}}\int_0^\varepsilon
\frac{\check{\rho}(\psi)d\psi}{\sqrt{\varepsilon-\psi}}.
\label{evendfpjse}
\end{equation}
It can be also found from (\ref{dfaqt}) that the two-integral even DF is of the form  
\begin{equation}
\tilde{f}_+(\widehat{Q})=
\frac{H(\widehat{Q})}{\sqrt{8}\pi^2}\frac{d^{2}}{d\widehat{Q}^{2}}\int_0^{\widehat{Q}}
\frac{\tilde{\rho}(\psi)d\psi}{\sqrt{\widehat{Q}-\psi}}
\label{evendfpjt}
\end{equation}
corresponding to the second term on the right side of (\ref{rhopj2}), 
where and below, $\widehat{Q}=\varepsilon-L_z^2/(2a^2).$ 

We can first know from (\ref{rhopj2htb}) that $\hat{\rho}(\psi)$ can 
be expanded as a power series of the form 
 \begin{equation}
\hat{\rho}(\psi)=\sum\limits_{n=5}^{+\infty}\hat{a}_n\psi^n. 
\label{rhoh-series}
\end{equation}
Then, by using (\ref{dfaq2}), the even DF corresponding to the third term on the right side of (\ref{rhopj2}) can be given by 
\begin{eqnarray}
\hat{f}_+(\varepsilon, L_z)=\left\{\begin{array}{cc}
\sum\limits_{n=5}^{+\infty}
\frac{\hat{a}_n\Gamma(n+1)\varepsilon^{n-3/2}}{(2\pi)^{3/2}\Gamma(n-1/2)}
\hbox{ }_2F_1\left(1,\frac{3}{2}-n;\frac{1}{2};\frac{L_z^2}{2\varepsilon a^2}\right) & {\rm as~~~} \varepsilon\geq L_z^2/(2a^2),  \\
-\sum\limits_{n=5}^{+\infty}
\frac{\hat{a}_n\Gamma(n+1)\varepsilon^{n-3/2}}{2(2\pi)^{3/2}\Gamma(n+1/2)}
\left(\frac{L_z^2}{2\varepsilon a^2}\right)^{-1}
\hbox{}_2F_1\left(1,\frac{3}{2};n+\frac{1}{2};\frac{2\varepsilon a^2}{L_z^2}\right) & {\rm as~~~} \varepsilon\leq L_z^2/(2a^2).
\end{array}\right.
\label{dfaqpj2}
\end{eqnarray} 
Note that the power series (\ref{rhoh-series})
converges so fast that we can use a sum of $N$ previous terms of (\ref{dfaqpj2}) as 
a viable approximation of the even DF, where  
 $N$ is a natural number such that $|\hat{a}_{N+5}|$ is less than a small accuracy given in numerical calculation. 
Then this approximation, denoted by  
$\hat{S}_+^N(\varepsilon, L_z)$,  can be explicitly given by 
\begin{eqnarray}
\hat{S}_+^N(\varepsilon, L_z)=\left\{\begin{array}{cc}
\sum\limits_{n=5}^{N+4}
\frac{\hat{a}_n\Gamma(n+1)\varepsilon^{n-3/2}}{(2\pi)^{3/2}\Gamma(n-1/2)}
\hbox{ }_2F_1\left(1,\frac{3}{2}-n;\frac{1}{2};\frac{L_z^2}{2\varepsilon a^2}\right) & {\rm as~~~} \varepsilon\geq L_z^2/(2a^2),  \\
-\sum\limits_{n=5}^{N+4}
\frac{\hat{a}_n\Gamma(n+1)\varepsilon^{n-3/2}}{2(2\pi)^{3/2}\Gamma(n+1/2)}
\left(\frac{L_z^2}{2\varepsilon a^2}\right)^{-1}
\hbox{}_2F_1\left(1,\frac{3}{2};n+\frac{1}{2};\frac{2\varepsilon a^2}{L_z^2}\right) & {\rm as~~~} \varepsilon\leq L_z^2/(2a^2).
\end{array}\right.
\label{dfaqpj2sum}
\end{eqnarray} 
Hence the above series expression is computationally superior to the contour integral. 

By using (\ref{dfaqb}), corresponding to the last term on the right side of (\ref{rhopj2}), 
the two-integral even DF $\bar{f}_+(\widehat{Q},L_z)$  
can be expressed as follows: 
\begin{equation}
\bar{f}_+(\widehat{Q},L_z)=
\frac{H(\widehat{Q})}{\sqrt{8}\pi^2}\left[\frac{d^{2}}{d\widehat{Q}^{2}}\int_0^{\widehat{Q}}
\frac{\bar{\rho}(\psi)d\psi}{\sqrt{\widehat{Q}-\psi}}
- \frac{L_z^2}{a^2} \frac{d^{3}}{d\widehat{Q}^{3}}\int_0^{\widehat{Q}}
\frac{\bar{\rho}(\psi)d\psi}{\sqrt{\widehat{Q}-\psi}}\right].
\label{evendfpjb}
\end{equation} 

Finally, by adding
all the above four DFs (\ref{evendfpjse}), (\ref{evendfpjt}), (\ref{dfaqpj2}) and (\ref{evendfpjb}), 
the two-integral even DF 
of the prolate Jaffe model with the mass density (\ref{rhopj2})  
can be obtained as a function 
$f_+(\varepsilon,L_z)=\check{f}_+(\varepsilon)+\tilde{f}_+(\widehat{Q})+\hat{f}_+(\varepsilon,L_z)+\bar{f}_+(\widehat{Q},L_z).$ Thus, 
when $\varepsilon> L_z^2/(2a^2),$ $f_+(\varepsilon,L_z)$ is given by 
\begin{eqnarray}
f_+(\varepsilon,L_z)
=\frac{H(\varepsilon)}{\sqrt{8}\pi^2}\frac{d^{2}}{d\varepsilon^{2}}\int_0^\varepsilon
\frac{\check{\rho}(\psi)d\psi}{\sqrt{\varepsilon-\psi}} +
\frac{1}{\sqrt{8}\pi^2}\frac{d^{2}}{d\widehat{Q}^{2}}\int_0^{\widehat{Q}}
\frac{\tilde{\rho}(\psi)d\psi}{\sqrt{\widehat{Q}-\psi}} \hspace*{7cm}\nonumber \\
+\sum\limits_{n=5}^{+\infty}
\frac{\hat{a}_n\Gamma(n+1)\varepsilon^{n-3/2}}{(2\pi)^{3/2}\Gamma(n-1/2)}
\hbox{ }_2F_1\left(1,\frac{3}{2}-n;\frac{1}{2};\frac{L_z^2}{2\varepsilon a^2}\right) %\nonumber \\
+\frac{1}{\sqrt{8}\pi^2}\left[\frac{d^{2}}{d\widehat{Q}^{2}}\int_0^{\widehat{Q}}
\frac{\bar{\rho}(\psi)d\psi}{\sqrt{\widehat{Q}-\psi}}
- \frac{L_z^2}{a^2} \frac{d^{3}}{d\widehat{Q}^{3}}\int_0^{\widehat{Q}}
\frac{\bar{\rho}(\psi)d\psi}{\sqrt{\widehat{Q}-\psi}}\right];
\label{evendfpj2-1}
\end{eqnarray}
when $\varepsilon\leq L_z^2/(2a^2),$ $f_+(\varepsilon,L_z)$ is of the form 
\begin{eqnarray}
f_+(\varepsilon,L_z)
=\frac{H(\varepsilon)}{\sqrt{8}\pi^2}\frac{d^{2}}{d\varepsilon^{2}}\int_0^\varepsilon
\frac{\check{\rho}(\psi)d\psi}{\sqrt{\varepsilon-\psi}} 
-\sum\limits_{n=5}^{+\infty}
\frac{\hat{a}_n\Gamma(n+1)\varepsilon^{n-3/2}}{2(2\pi)^{3/2}\Gamma(n+1/2)}
\left(\frac{L_z^2}{2\varepsilon a^2}\right)^{-1}
\hbox{}_2F_1\left(1,\frac{3}{2};n+\frac{1}{2};\frac{2\varepsilon a^2}{L_z^2}\right). 
\label{evendfpj2-2}
\end{eqnarray}

Now let us consider the odd part of the DF for the prolate Jaffe model under the assumption of the rotation law. 
It can be known from (\ref{rhopjs}) and (\ref{rhopj2htb}) that the functions 
$\check{\rho}(\psi),$ $\tilde{\rho}(\psi)$ and $\bar{\rho}(\psi)$ can be expanded as three power series: 
 \begin{equation}
\check{\rho}(\psi)=\sum\limits_{n=4}^{+\infty}\check{a}_n\psi^n, ~~~
\tilde{\rho}(\psi)=\sum\limits_{n=3}^{+\infty}\tilde{a}_n\psi^n,  ~~~
\bar{\rho}(\psi)=\sum\limits_{n=3}^{+\infty}\bar{a}_n\psi^n. 
\label{rhostb-series}
\end{equation}
Similarly, by first inserting  (\ref{rhoh-series}) and (\ref{rhostb-series}) into (\ref{rhopj2}) and then using
(\ref{dfaqoddall}), we can obtain the two-integral odd DF   
\begin{eqnarray}
f_\_(\varepsilon,L_z)=\frac{a^{2k}{\rm sgn}(L_z)}{4\pi\Gamma(k)}
\sum\limits_{n=4}^{+\infty}
\check{a}_n n!\varepsilon^{n-2}
{\mathbb H}\left(k,0,n-1,1;\frac{L_z^2}{2\varepsilon a^2}\right) %\nonumber\\
+\frac{a^{2k-1}{\rm sgn}(L_z)}{4\pi\Gamma(k+1/2)}
\sum\limits_{n=3}^{+\infty}
\tilde{a}_n n!\varepsilon^{n-2}
{\mathbb H}\left(k,\frac{1}{2},n-1,1;\frac{L_z^2}{2\varepsilon a^2}\right) \nonumber\\
+\frac{a^{2k-2}{\rm sgn}(L_z)}{4\pi\Gamma(k+1)}
\sum\limits_{n=5}^{+\infty}
\hat{a}_n n!\varepsilon^{n-2}
{\mathbb H}\left(k,1,n-1,1;\frac{L_z^2}{2\varepsilon a^2}\right)  %\nonumber\\
+\frac{a^{2k-3}{\rm sgn}(L_z)}{4\pi\Gamma(k+3/2)}
\sum\limits_{n=3}^{+\infty}
\bar{a}_n n!\varepsilon^{n-2}
{\mathbb H}\left(k,\frac{3}{2},n-1,1;\frac{L_z^2}{2\varepsilon a^2}\right)
\label{odddfpj}
\end{eqnarray}
for the prolate Jaffe model with the mass density (\ref{rhopj2}),  
constrained by the rotation law of the form  
$\langle{v}_\phi\rangle=R^{2k}/(1+R^2/a^2)^k$ 
for $k>0.$ 
In particular, when $0\leq [k]\leq 2,$ it can be deduced 
from (\ref{dfaqodd})  that the two-integral odd DF   
corresponding to the third term on the right side of (\ref{rhopj2}) 
 can be given by 
\begin{equation}
\hat{f}_\_(\widehat{Q},L_z)=\frac{(k+1){\rm sgn}(L_z)L_z^{2k}H(\widehat{Q})}{4\pi 2^ka^2\Gamma(k+2)\Gamma(1-k_0)}
\frac{d^{[k]+3}}{d\widehat{Q}^{[k]+3}}\int_0^{\widehat{Q}}  
\frac{\hat{\rho}(\psi)d\psi}{(\widehat{Q}-\psi)^{k_0}}. 
\label{odddfpjh}
\end{equation}
Hence, by (\ref{dfaqoddall}) and (\ref{odddfpjh}), we can also obtain the odd DF 
\begin{eqnarray}
f_\_(\varepsilon,\widehat{Q},L_z)=\frac{a^{2k}{\rm sgn}(L_z)}{4\pi\Gamma(k)}
\sum\limits_{n=4}^{+\infty}
\check{a}_n n!\varepsilon^{n-2}
{\mathbb H}\left(k,0,n-1,1;\frac{L_z^2}{2\varepsilon a^2}\right) %\nonumber\\
+\frac{a^{2k-1}{\rm sgn}(L_z)}{4\pi\Gamma(k+1/2)}
\sum\limits_{n=3}^{+\infty}
\tilde{a}_n n!\varepsilon^{n-2}
{\mathbb H}\left(k,\frac{1}{2},n-1,1;\frac{L_z^2}{2\varepsilon a^2}\right) \nonumber\\
+\frac{(k+1){\rm sgn}(L_z)L_z^{2k}H(\widehat{Q})}{4\pi 2^ka^2\Gamma(k+2)\Gamma(1-k_0)}
\frac{d^{[k]+3}}{d\widehat{Q}^{[k]+3}}\int_0^{\widehat{Q}}  
\frac{\hat{\rho}(\psi)d\psi}{(\widehat{Q}-\psi)^{k_0}} %\hspace*{2.5cm}\nonumber\\
+\frac{a^{2k-3}{\rm sgn}(L_z)}{4\pi\Gamma(k+3/2)}
\sum\limits_{n=3}^{+\infty}
\bar{a}_n n!\varepsilon^{n-2}
{\mathbb H}\left(k,\frac{3}{2},n-1,1;\frac{L_z^2}{2\varepsilon a^2}\right)
\label{odddfpj2}
\end{eqnarray}
for the prolate Jaffe model with the mass density (\ref{rhopj2}) 
under the above rotation law for $[k]\in (0,2].$

All the above series are theoretically convergent to the corresponding 
Hunter-Qian contour integrals, that is, 
these series expressions of the DFs can be in fact used 
to recover Hunter and Qian's formulae of the DFs for the prolate models. 
The convergence of these series is fast enough that we can adopt 
 a sum of finite previous terms of series as a viable approximation of the DFs 
for understanding of dynamical properties of the stellar system. 
Let $\delta_0$ denote the desired small accuracy. 
Then this approximation of the DF (\ref{odddfpj2}), denoted by $S_\_^l(\varepsilon,\widehat{Q},L_z),$ 
can be given by
\begin{eqnarray}
S_\_^l(\varepsilon,\widehat{Q},L_z)=\frac{a^{2k}{\rm sgn}(L_z)}{4\pi\Gamma(k)}
\sum\limits_{n=4}^{l+3}
\check{a}_n n!\varepsilon^{n-2}
{\mathbb H}\left(k,0,n-1,1;\frac{L_z^2}{2\varepsilon a^2}\right) %\nonumber\\
+\frac{a^{2k-1}{\rm sgn}(L_z)}{4\pi\Gamma(k+1/2)}
\sum\limits_{n=3}^{l+2}
\tilde{a}_n n!\varepsilon^{n-2}
{\mathbb H}\left(k,\frac{1}{2},n-1,1;\frac{L_z^2}{2\varepsilon a^2}\right) \nonumber\\
+\frac{(k+1){\rm sgn}(L_z)L_z^{2k}H(\widehat{Q})}{4\pi 2^ka^2\Gamma(k+2)\Gamma(1-k_0)}
\frac{d^{[k]+3}}{d\widehat{Q}^{[k]+3}}\int_0^{\widehat{Q}}  
\frac{\hat{\rho}(\psi)d\psi}{(\widehat{Q}-\psi)^{k_0}} %\hspace*{2.5cm}\nonumber\\
+\frac{a^{2k-3}{\rm sgn}(L_z)}{4\pi\Gamma(k+3/2)}
\sum\limits_{n=3}^{l+2}
\bar{a}_n n!\varepsilon^{n-2}
{\mathbb H}\left(k,\frac{3}{2},n-1,1;\frac{L_z^2}{2\varepsilon a^2}\right).
\label{odddfpj2sum}
\end{eqnarray}
where $l$ is dependent of the convergence of the power series 
given by (\ref{rhostb-series}), for example, 
we can choose  the natural number $l$  to be 
such that $|\check{a}_{l+4}|+|\tilde{a}_{l+3}|+|\bar{a}_{l+3}|<\delta_0$ 
since we find that the convergence of these power series is fast enough. 
Fortunately, all the terms in these series can be expressed 
in terms of hypergeometric functions and 
it is easy to calculate numerically these hypergeometric functions. Therefore  
this approximation method is in general computationally preferable to the contour integral method, 
especially for the prolate Plummer model considered below. 

Of course,  for the prolate Jaffe models, 
Jiang and Moss (2002) gave the complex contour integral expressions 
of the even DFs and obtained explicit formulae of both velocity dispersions and rotation curves.
We can know from Hunter's (1977) formulae that these velocity dispersions for the prolate models 
can be expressed in terms of elementary functions of the two variables $R$ and $z.$  
We can also find that 
the prolate models have anisotropic velocity distributions. 
More other properties of the prolate models have been given by Jiang and Moss (2002) 
including some figures such as the even DFs, velocity dispersions and rotational velocity. 
Their details are therefore not repeated here. 

It is finally worth mentioning that a different way to recover Hunter and Qian's formulae of the DFs for the prolate models
  is to employ the Laplace-Mellin transformation given in Section \ref{even}.  

\subsection{The prolate Plummer model}
\label{edfplummer}
The gravitational potential of the prolate Plummer model is given by 
\begin{equation}
\Phi(R,z)=-\frac{GM}{\sqrt{(\sqrt{R^2+a^2}+b)^2+z^2}}
\label{Phipp}
\end{equation}
where $a,$ $b,$ $M$ and $G$ are the same as in (\ref{Phipj}).  As $b$ goes to zero, 
(\ref{Phipp}) becomes a potential of the spherical Plummer (1911) model.   
The mass density corresponding to (\ref{Phipp}) can be calculated as 
\begin{equation}
\rho(R,z)=\frac{M}{4\pi}
\frac{b\tau^2(X^2+a^2) +3a^2X(X+b)^2}
{\tau^5X^3}
\label{rhopp}
\end{equation}
where $X$ and $\tau$ are the same as in (\ref{rhopj}). 
It can be found that the model defined by (\ref{Phipp}) and (\ref{rhopp}) is prolate 
and that for any given $a>0,$ the larger the value of $b,$ the more prolate this model. 

Similarly, by letting  
$\psi=-\Phi(R,z)$ and using (\ref{Phipp}), the mass density (\ref{rhopp}) can be rewritten  
as a function $\rho(\psi,R)$ of $\psi$ and $R,$ that is, it is of the following form
\begin{equation}
\rho(\psi,R)=\frac{3a^2\psi^5}{4\pi G^5M^4}
+\left[\frac{6a^2b\psi^5}{4\pi G^5M^4}+\frac{b\psi^3}{4\pi G^3M^2}\right]
\frac{1}{X} 
+\frac{3a^2b^2\psi^5}{4\pi G^5M^4}\frac{1}{X^2}
+\frac{b\psi^3}{4\pi G^3M^2}\frac{1}{X^3}.
\label{rhopp2}
\end{equation} 
Take $\widehat{Q}=\varepsilon- L_z^2/(2a^2).$ 
Then, by using (\ref{dfaq2}), (\ref{dfaqt}), (\ref{dfaqb}) and (\ref{dfs}), 
 the two-integral even DF $f_+(\varepsilon,L_z)$ corresponding to the mass density (\ref{rhopp2}) can be obtained as follows.  
When $\widehat{Q}\leq 0,$
\begin{eqnarray}
f_+(\varepsilon,L_z)= 
\frac{96a^2\varepsilon^{7/2}H(\varepsilon)}{7\sqrt{8}\pi^3G^5M^4} 
-\frac{64a^2b^2\varepsilon^{9/2}}{21\sqrt{8}\pi^3 G^5M^4L_z^2}
\hbox{ }_2F_1\left(1,\frac{3}{2};\frac{11}{2};\frac{2\varepsilon a^2}{L_z^2}\right); 
\label{evendfpp-1}
\end{eqnarray}
when $\widehat{Q}> 0,$
\begin{eqnarray}
f_+(\varepsilon,L_z)= 
\frac{96a^2\varepsilon^{7/2}H(\varepsilon)}{7\sqrt{8}\pi^3G^5M^4} +
\frac{2b\widehat{Q}^{3/2}(96 a^2\widehat{Q}^{2}+7G^2M^2)H(\widehat{Q})}{7\sqrt{8}\pi^3aG^5M^4}\hspace*{4cm} \nonumber \\
+\frac{96b^2\varepsilon^{7/2}}{7\sqrt{8}\pi^3 G^5M^4}
\hbox{ }_2F_1\left(1,-\frac{7}{2};\frac{1}{2};\frac{L_z^2}{2\varepsilon a^2}\right)
+\frac{b\widehat{Q}^{1/2}(2a^2\widehat{Q}-3L_z^2)H(\widehat{Q})}{\sqrt{8}\pi^3a^5G^3M^2}. 
\label{evendfpp-2}
\end{eqnarray}
Obviously, it is easier to calculate numerically 
the even DF given by (\ref{evendfpp-1}) and (\ref{evendfpp-2})  than the Hunter-Qian complex integral. 

 Similarly, we can also give a simple expression of the odd DF for the prolate Plummer model 
constrained by the rotation curve. For example, if the rotation curve is assumed to be 
determined by $\langle{v}_\phi\rangle=R^{2}/(1+R^2/a^2),$ then, by (\ref{dfaqoddall}), 
the odd part of the DF corresponding to the mass density (\ref{rhopp2}) can be 
obtained as 
\begin{eqnarray}
f_\_(\varepsilon,L_z)=a^2{\rm sgn}(L_z)\left[
\frac{45a^2\varepsilon^{3}}{2\pi^2 G^5M^4}
{\mathbb H}\left(1,0,3,1;\frac{L_z^2}{2\varepsilon a^2}\right)
+\frac{90a^2b\varepsilon^{3}}{\pi^{5/2} G^5M^4}
{\mathbb H}\left(1,\frac{1}{2},3,1;\frac{L_z^2}{2\varepsilon a^2}\right) \right.  \hspace*{3.5cm} \nonumber \\ \left.
+\frac{3b\varepsilon}{4\pi^{5/2} G^3M^2}
{\mathbb H}\left(1,\frac{1}{2},2,1;\frac{L_z^2}{2\varepsilon a^2}\right) 
+\frac{45a^2b^2\varepsilon^{3}}{2\pi^2 G^5M^4}
{\mathbb H}\left(1,1,3,1;\frac{L_z^2}{2\varepsilon a^2}\right)
+\frac{b\varepsilon}{2\pi^{5/2} G^3M^2}
{\mathbb H}\left(1,\frac{3}{2},2,1;\frac{L_z^2}{2\varepsilon a^2}\right)
\right].
 \label{dfaqoddpp}
\end{eqnarray} 
All the functions ${\mathbb H}$ presented here can be expressed in terms of hypergeometric functions 
which can be calculated numerically with little additional effort. 

\section{Conclusions}
\label{con}
It is recently shown that near the centre the potentials of the dark matter halos are 
approximately prolate on average (Hayashi, Navarro and Springel 2007). 
This implies that the study of prolate models   
is very significant for the investigation of the dark matter halos of the galaxies. 
To do this, it is a natural idea to explore different expressions of the two-integral DFs 
in axisymmetric stellar systems.
These expressions presented here can be successfully applied into 
getting the even and odd parts of the  two-integral DFs for both the prolate Jaffe and Plummer models. 
We can thus show that both the even and odd parts of the two-integral DFs for 
the prolate Jaffe model can be expressed as series 
 convergent to  the corresponding contour integrals obtained by use of Hunter and Qian's (1993) contour integral formulae.  
The convergence of these series is fast enough that we can 
use a sum of finite previous terms of series as a viable approximation of the two-integral DFs. 
All the terms of series can be expressed in terms of hypergeometric functions 
and it is easy to calculate numerically these hypergeometric functions. 
 Hence this method is generally superior to 
the Hunter-Qian contour method, especially for the prolate Plummer model. 
Similar such expressions of series for the two-integral DFs 
have been considered for some stellar systems (e.g. Miyamoto 1971).  

It is remarkable that the two-integral even DFs for some separable densities such as (\ref{rho1t}) and (\ref{rho1b}) 
can be expressed as a sum of finite real integrals which are easier to calculate numerically 
than their corresponding complex contour integrals. 
We can also find that the Hunter-Qian contour integral 
formula of the two-integral even DF for axisymmetric systems can be recovered 
by use of the Laplace-Mellin integral transformation originally developed by Dejonghe (1986). 

According to the Hunter (1977) formulae, a simple expression of the radial,  vertical and rotational velocity dispersions 
 can be directly given by using the axisymmetric potentials in the stellar systems. 
It can be also known that the radial velocity dispersion  
is equal to the vertical velocity dispersion for all the two-integral models in axisymmetric stellar systems. 
However, in real axisymmetric stellar systems, the velocity dispersion in the radial direction 
is not equal to the velocity dispersion in the vertical direction. This means that the DFs of the real systems 
must actually depend on three integrals of the motion. In axisymmetric systems, there are two classical  
isolating integrals, that is,  the energy and the angular momentum with respect to the axis of symmetry.     
It can be also found by numerical calculation that the third isolating integral is respected by orbits in 
realistic axisymmetric galactic potentials. The third isolating integral 
cannot be generally expressed by using basic elementary functions of energy and angular momenta.  But 
some three-integral models have been constructed  for particular orbital families in flattened axisymmetric systems 
(Evans et {al.} 1997) and for separable axisymmetric St\"{a}ckel potentials (Famaey et {al.} 2002). 

 It is hence a very significant research field in both mathematics and astrophysics 
to consider the three-integral DFs for a good description of elliptical galaxies. 
Any three-integral model requires the existence of a global third isolating integral 
and its study is now still based on special potentials which allow the three-integral DFs 
to be obtained as tractable expressions. For example, Dejonghe and de Zeeuw (1988) first constructed 
the three-integral DFs for the Kuzmin-Kutuzov (1962) models since these models have St\"ackel's potentials. 
In general, either the prolate Jaffe potential or the prolate Plummer potential is not of the St\"ackel type 
that seems the most general of integrable potentials (Lynden-Bell 1962b). 
The question addressed here is, is it possible to develop the three-integral DFs 
either for the prolate Jaffe model or for the prolate Plummer potential? 
This is an open problem now. 
To answer this question, we have first to see whether there exists a global third isolating integral for the 
prolate Jaffe model, which can be expressed in terms of energy and angular momenta. 
Once the third integral is found, a tractable expression of the three-integral DFs can 
be given as Dejonghe and de Zeeuw (1988) did. 
In principle, this is a reasonably clear procedure. However, so far it has only been carried through 
for special cases mentioned above; it seems difficult to obtain useful results for more general cases. 

\vskip0.2cm
\noindent {\bf Acknowledgment}.  
The first author was supported by NSFC 10271121 and by 
SRF for ROCS, SEM.  
The second author was supported by Leading Scientific School grant 1078.2003.02. 
The cooperation of authors was supported by joint grants 
of NSFC 10511120278/10611120371 and RFBR 04-02-39026. 
The two authors are very grateful to Professor Konstantin Kholshevnikov, 
 Sergei Kutuzov and Vadim Antonov 
for their valuable discussions in this work.  
The two authors would also like to thank the referee of this paper 
for his/her valuable comments and suggestions in this work.

\bsp
\label{lastpage}
\end{document}